\documentclass[aps,prb,showpacs,twocolumn,superscriptaddress]{revtex4} 
\usepackage{bm,color}
\usepackage[dvipdfmx]{graphicx}
\usepackage{amsmath}
\usepackage{color}
\usepackage{here}
\begin{document}
\title {Theoretical study on stabilization and destabilization of magnetic skyrmions by uniaxial-strain-induced anisotropic Dzyaloshinskii--Moriya interactions}
\author{Kohei Tanaka}
\affiliation{Department of Nano Science and Engineering, Waseda University, Okubo, Shinjuku-ku, Tokyo 169-8555, Japan}
\author{Ryosuke Sugawara}
\affiliation{Department of Applied Physics, Waseda University, Okubo, Shinjuku-ku, Tokyo 169-8555, Japan}
\author{Masahito Mochizuki}
\affiliation{Department of Applied Physics, Waseda University, Okubo, Shinjuku-ku, Tokyo 169-8555, Japan}
\begin{abstract}
Magnetic skyrmions in chiral-lattice ferromagnets are currently attracting enormous research interest because of their potential applications in spintronic devices. However, they emerge in bulk specimens only in a narrow window of temperature and magnetic field. This limited stability regime is recognized as an obstacle to technical applications. Recent experiments demonstrated that the thermodynamic stability of magnetic skyrmions is enhanced or suppressed by the application of a uniaxial strain depending on its axial direction in bulk chiral-lattice ferromagnets MnSi [Y. Nii $et$ $al.$, Nat. Commun. {\bf 6}, 8539 (2015), A. Chacon $et$ $al.$, Phys. Rev. Lett. {\bf 115}, 267202 (2015)] and Cu$_2$OSeO$_3$ [S. Seki $et$ $al.$, Phys. Rev. B {\bf 96}, 220404(R) (2017)]. Motivated by these experimental discoveries, we theoretically investigated the effects of anisotropic Dzyaloshinskii--Moriya interactions on the stability of magnetic skyrmions caused by this uniaxial strain. We find that magnetic skyrmions are significantly stabilized (destabilized) in the presence of anisotropic DM interactions when an external magnetic field lies perpendicular (parallel) to the anisotropy axis, along which the DM coupling is strengthened. Our results account completely for the experimentally observed strain-induced stabilization and destabilization of magnetic skyrmions and provide a firm ground for possible strain engineering of skyrmion-based electronic devices.
\end{abstract}
\maketitle

\section{Introduction}
Keen competition between the Dzyaloshinskii--Moriya (DM) interactions~\cite{Dzyaloshinsky58,Moriya60} and the ferromagnetic exchange interactions in chiral-lattice ferromagnets often results in the formation of magnetic skyrmions~\cite{Bogdanov89,Bogdanov99,Rossler06}, that is, vortex-like nanometric spin textures characterized by a quantized topological invariant~\cite{Nagaosa13,Everschor18}. Their realization was experimentally discovered in metallic ferromagnets MnSi~\cite{Muhlbauer09,Tonomura12} and Fe$_{1-x}$Co$_x$Si~\cite{YuXZ10,Munzer10}, which have a chiral cubic crystalline structure. Immediately after these discoveries, it was revealed that the magnetic skyrmions can be driven or manipulated with ultralow electric-current densities~\cite{Jonietz10,YuXZ12,Schulz12}. The threshold current density turned out to be five or six orders of magnitude smaller than that required to drive ferromagnetic domain walls~\cite{Yamanouchi04,Tatara04,Barnes05}. Subsequent theoretical work based on Thiele's equation found that this high mobility of magnetic skyrmions is attributable to their topological nature~\cite{Everschor12,Iwasaki13a,Iwasaki13b}.

Because of their nanometric size and high mobility, magnetic skyrmions are recognized as potential information carriers in future magnetic storage devices of high information density and low energy consumption~\cite{Fert13,Tomasello14,Koshibae15}. Moreover, intensive studies have uncovered their numerous functionalities, and they are now recognized as promising building blocks of versatile functional devices~\cite{Finocchio16}, e.g., logic gates~\cite{ZhangX15}, microwave detection/generation~\cite{Finocchio15}, and brain-inspired computations~\cite{Pinna18b,Prychynenko18,Bourianoff18}.

However, the skyrmion phase in these magnets is known to be thermodynamically unstable and only appears in a tiny window of temperature $T$ and magnetic field $H$ below a magnetic ordering temperature in the phase diagram~\cite{Muhlbauer09,Neubauer09}. Indeed, since the discovery of magnetic skyrmions in MnSi~\cite{Muhlbauer09} and Fe$_{1-x}$Co$_x$Si~\cite{YuXZ10,Munzer10}, many skyrmion-hosting materials have been discovered~\cite{Wilhelm11,Seki12a,Seki12b,Seki12c,Adams12,Kezsmarki15,Tokunaga15}, and all these compounds turned out to exhibit similar $T$--$H$ phase diagrams with a very tiny skyrmion phase regime despite the different crystalline structures and distinct origins of magnetism.

This limited stability of magnetic skyrmions is recognized as an obstacle to technical applications. Therefore, a lot of experimental efforts have been devoted to enhance their stability. Yu and coworkers discovered that the skyrmion phase is strongly stabilized in thin samples, the thickness of which is comparable or thinner than the magnetic modulation period~\cite{YuXZ11}. Subsequent theoretical work has accounted for this phenomena~\cite{Butenko10}. However, this method restricts the sample shape to thin films or thin plates and thus is not applicable to arbitrary shapes of the sample. It was also reported that applications of hydrostatic pressures~\cite{Ritz13,Levatic16} and electric fields~\cite{Okamura16} may enhance the stability of skyrmions slightly. However, the induced changes in temperature range turned out to be very tiny (only by several degrees Kelvin). Another interesting experiment is the rapid cooling of the sample, which often gives rise to a supercooled skyrmion crystal phase that spreads widely in the $T$--$H$ phase diagram~\cite{Okamura16,Oike16,Karube16}. However, this phase is not a thermal equilibrium phase but a metastable state with a finite lifetime. Therefore, more efficient and elaborate methods to realize these thermodynamically stable skyrmions have been eagerly awaited.

Under these circumstances, experimental applications of uniaxial compressive strain was found to stabilize or destabilize the magnetic skyrmions in MnSi~\cite{Nii15,Chacon15} depending on the relative direction of the uniaxial strain against the external $\bm H$ field. They observed that the skyrmion crystal phase regime in the $T$--$H$ phase diagram for MnSi expands (shrinks) when a uniaxial strain is applied perpendicular (parallel) to the external magnetic field $\bm H$. Subsequently, Seki and collaborators reported dramatic changes in stability of magnetic skyrmions by application of a uniaxial strain to chiral-lattice ferrimagnetic insulator Cu$_2$OSeO$_3$~\cite{Seki17}. They discovered that the uniaxial strain applied perpendicular to $\bm H$ again widens the skyrmion crystal phase regime significantly, whereas the strain applied parallel to $\bm H$ destabilizes it resulting in the disappearance of the skyrmion crystal phase. 

In these experiments, it is expected that the uniaxial compressive strain strengthens the DM coupling on compressed bonds in MnSi, whereas the uniaxial strain strengthens the DM coupling on stretched bonds in Cu$_2$OSeO$_3$ by enhancing the spatial inversion asymmetry of their crystallographic structures. This uniaxial enhancement of the DM coupling in the presence of uniaxial crystalline distortion has been confirmed in experiments on a chiral-lattice ferromagnet FeGe~\cite{Shibata15}. Note that whether bond compression or bond stretching strengthens the DM coupling may depend on details of the electronic and crystal structures in materials. The variation of DM coupling upon the uniaxial crystallographic distortion as well as their quantitative evaluations may require more microscopic studies based on the first-principles calculations~\cite{Koretsune15,Koretsune18}.

Motivated by these experimental findings, we theoretically study the effects of anisotropic DM interactions on the stability of magnetic skyrmions caused by a uniaxial strain in bulk chiral-lattice ferromagnets based on numerical analyses of a classical spin model. We show that the anisotropic DM coupling indeed stabilizes or destabilizes the skyrmion crystal phase depending on the relative directional combinations of uniaxial strain and external magnetic field $\bm H$. The skyrmion crystal phase regime spreads even to the lowest temperature in the $T$--$H$ phase diagram when $\bm H$ is applied perpendicular to the uniaxial strain or the anisotropy axis along which the DM coupling is strengthened. Conversely, this regime shrinks or even vanishes when $\bm H$ is applied parallel to the uniaxial strain. These results thoroughly account for the observed strain-induced stabilization and destabilization of magnetic skyrmions in experiments. Our work provides a firm basis to possible strain engineering of magnetic skyrmions towards future skyrmion-based spin electronics.

\section{Model and Methods}
To describe the magnetism in a bulk chiral-lattice ferromagnet, we start with the classical Heisenberg model on a cubic lattice~\cite{Bak80}:
\begin{eqnarray}
\mathcal H_1&=&
-J\sum_{i,\bm \hat{\bm \gamma}} {\bm m}_i\cdot{\bm m}_{i+\hat{\bm \gamma}}
- \sum_{i,\bm \hat{\bm \gamma}} D_\gamma ({\bm m}_i\times{\bm m}_{i+\hat{\bm \gamma}}
\cdot\hat{\bm \gamma})
\nonumber \\ & &
-\bm H \cdot \sum_i \bm m_i
\label{eq:H1}
\end{eqnarray}
where $\bm m_i$ represents a normalized magnetization vector, and $\hat{\bm \gamma}(=\hat{\bm x}, \hat{\bm y}, \hat{\bm z})$ is a unit directional vector pointing in the $\gamma(=x, y, z)$ direction. The first and second terms describe the ferromagnetic exchange interactions and the DM interactions, respectively, for the nearest-neighbor magnetization pairs, where $J$ and $D_\gamma$ denote their coupling coefficients. We use the three DM parameters $D_x$, $D_y$, and $D_z$ to treat the anisotropic DM coupling induced by the uniaxial strain. The application of uniaxial strain in the $\gamma$ direction is taken into account by increasing the DM parameter $D_\gamma$. The last term describes the Zeeman interaction associated with an external magnetic field $\bm H=(0,0,H)$ applied in the $z$ direction.

Note that, in the present work, we consider that all magnetic structures varying slowly in space, and, thereby, their coupling to the background crystalline structure is negligibly weak. This fact justifies our theoretical treatment based on the simple cubic lattice without considering the complicated real crystalline structure after a coarse graining of magnetization distributions and a division of space into cubic cells. However, as pointed out by Buhrandt and Fritz~\cite{Buhrandt13}, we need to take care of artificial magnetic anisotropies caused by this cell discretization. To consider these anisotropies, we rewrite the first and second terms of $\mathcal{H}_1$ using a Fourier transformation,
\begin{eqnarray}
& &\mathcal H_{\rm FM}
=\sum_{\bm k}\alpha_{\bm k} \bm m_{\bm k} \cdot \bm m_{-\bm k},
\\
& &\mathcal H_{\rm DM}=\sum_{\bm k} \beta_{\gamma \bm k}(\bm m_{\bm k} 
\times \bm m_{-\bm k}) \cdot \hat{\bm \gamma},
\end{eqnarray}
with
\begin{eqnarray}
\alpha_{\bm k}
&=&-J\left(\cos(k_xa)+\cos(k_ya)+\cos(k_za)\right) 
\nonumber \\
&=&-3J+\frac{a^2 J}{2}(k_x^2+k_y^2+k_z^2)-\frac{a^4J}{24}(k_x^4+k_y^4+k_z^4)
\nonumber \\
& &+{\mathcal O}(k^6),
\\
\beta_{\gamma \bm k}
&=&-D_\gamma \sin(k_\gamma a)
\nonumber \\
&=&-aD_\gamma k_\gamma + \frac{a^3 D_\gamma}{6}k_\gamma^3+\mathcal{O}(k^5)
\end{eqnarray}
where $a$ is the lattice constant of the cubic lattice.

The term $\displaystyle-\frac{a^4J}{24}(k_x^4+k_y^4+k_z^4)$ in $\alpha_{\bm k}$ and the term $\displaystyle \frac{a^3 D}{6}k_\gamma^3$ in $\beta_{\gamma \bm k}$ give deviations from spherical symmetry and eventually induce magnetic anisotropies. As argued in Ref.~\cite{Buhrandt13}, these artificial anisotropies may be compensated by involving the third nearest-neighbor interactions, which are given by
\begin{eqnarray}
\mathcal H_2
=J^{\prime}\sum_{i,\bm \hat{\bm \gamma}}{\bm m}_i
\cdot{\bm m}_{i+2\hat{\bm \gamma}}
+ \sum_{i,\bm \hat{\bm \gamma}} D^{\prime}_\gamma 
({\bm m}_i\times{\bm m}_{i+2\hat{\bm \gamma}}\cdot\hat{\bm \gamma}).
\label{eq:H2}
\end{eqnarray}
After a Fourier transformation of the total Hamiltonian $\mathcal{H}=\mathcal{H}_1+\mathcal{H}_2$, we obtain $\alpha_{\bm k}$ and $\beta_{{\gamma}\bm k}$,
\begin{eqnarray}
\alpha_{\bm k}&=&
-3(J-J^{\prime})+\frac{a^2}{2}(J-4J^\prime)(k_x^2+k_y^2+k_z^2)
\nonumber \\ & &
-\frac{a^4}{24}(J-16J^\prime)(k_x^4+k_y^4+k_z^4)+\mathcal{O}(k_\gamma^6)
\\
\beta_{\gamma \bm k}&=&
-a(D_\gamma-2D_\gamma^\prime)k_\gamma+\frac{a^3}{6}(D_\gamma-8D_\gamma^\prime)k_\gamma^3
\nonumber \\ & &
+\mathcal{O}(k^5)
\end{eqnarray}
From these expressions, we find that the artificial magnetic anisotropies vanish when
\begin{eqnarray}
J^\prime=\frac{1}{16}J, \quad\; D_\gamma^\prime=\frac{1}{8}D_\gamma,
\end{eqnarray}
as far up as fourth-order terms with respect to $k$ are concerned.

On the basis of the above discussion, we employ the following classical spin Hamiltonian,
\begin{eqnarray}
\mathcal H&=&
-J\sum_{i,\bm \hat{\bm \gamma}} {\bm m}_i\cdot{\bm m}_{i+\hat{\bm \gamma}}
+J^{\prime} \sum_{i,\bm \hat{\bm \gamma}} {\bm m}_i\cdot{\bm m}_{i+2\hat{\bm \gamma}}
\nonumber \\ & &
- \sum_{i,\bm \hat{\bm \gamma}} D_\gamma ({\bm m}_i
\times{\bm m}_{i+\hat{\bm \gamma}}
\cdot\hat{\bm \gamma})
\nonumber \\ & &
+ \sum_{i,\bm \hat{\bm \gamma}} D^{\prime}_\gamma 
({\bm m}_i\times{\bm m}_{i+2\hat{\bm \gamma}}\cdot 2\hat{\bm \gamma})
-H \sum_{i} m_{iz}
\label{eq:Htot}
\end{eqnarray}
In the following, we take $J=1$ as the unit of energy and consider a cubic lattice of $N$=$30 \times 30 \times 30$ sites with periodic boundary conditions. We study the ground-state properties at $T$=0 by minimizing the energies of various magnetic states by relaxing their spatial magnetization configurations. For this purpose, we first prepare initial magnetic configurations by performing a Monte Carlo thermalization at low temperatures, and further relax them by numerically solving the Landau--Lifshitz--Gilbert equation using the fourth-order Runge--Kutta method. In contrast, the thermodynamic properties at finite temperatures were studied using the Monte Carlo technique based on the Metropolis algorithm. We employed the replica-exchange Monte Carlo method as a measure for the system to avoid getting trapped in local energy minima~\cite{Swendsen86,Hukushima96,Earl05}.

\begin{figure*}[th]
\includegraphics[scale=0.45]{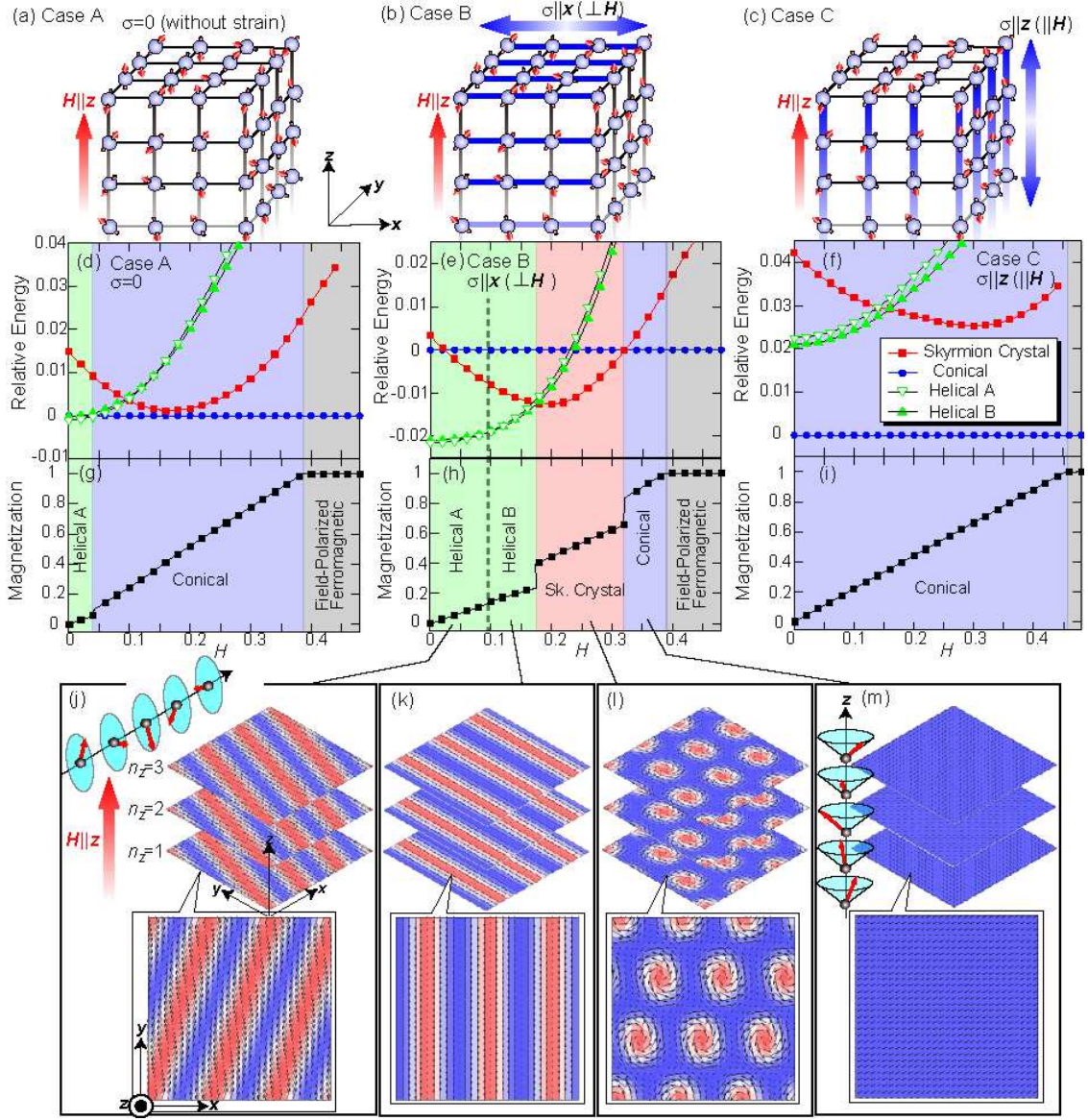}
\caption{(a)--(c) Three different cases were examined in the present study. The external magnetic field $\bm H$ was applied parallel to the $z$ axis for all cases. (a) Case A with isotropic DM interactions, $D_x=D_y=D_z$, which corresponds to a system without strain. (b) Case B with anisotropic DM interactions, $D_x>D_y=D_z$. This condition corresponds to a system to which a uniaxial strain $\bm \sigma$ perpendicular to the $\bm H$ field ($\bm \sigma$$\perp$$\bm H$) is applied in the $x$ direction. (c) Case C with anisotropic DM interactions, $D_z>D_x=D_y$. This condition corresponds to a system to which a uniaxial strain $\bm \sigma$ parallel to the $\bm H$ field ($\bm \sigma$$\parallel$$\bm H$) is applied in the $z$ direction. (d)--(f) Calculated $H$-profiles of relative energies of various magnetic states at $T$=0 for (d) Case A with $D_x$=$D_y$=$D_z$=0.727, (e) Case B with $D_x$=0.8 and $D_y$=$D_z$=0.727, and (f) Case C with $D_z$=0.8 and $D_x$=$D_y$=0.727. (g)--(i) Calculated $H$-profiles of net magnetization for (g) Case A, (h) Case B, and (i) Case C. From these $H$-profiles of relative energies and net magnetizations, we drew the phase diagrams as function of $H$ at $T=0$ for the three cases. (j)--(m) Magnetization configurations of examined magnetic states. (j) helical A, (k) helical B, (l) skyrmion crystal, and (m) conical states. The layered magnetic structures are stacked uniformly along the $z$ axis.}
\label{Fig1}
\end{figure*}
\section{Results}
We examined a case with isotropic DM interactions with $D_x$=$D_y$=$D_z$ (Case A) and two cases of anisotropic DM interactions with different axial anisotropy directions against the $\bm H$ field (Cases B and C) [see Fig.~\ref{Fig1}(a)--(c)], where $\bm H$ is always applied along the $z$ axis ($\bm H$$\parallel$$\bm z$). Case B corresponds to the anisotropic DM interactions with $D_x>D_y=D_z$, and Case C to those with $D_z>D_x=D_y$. Here, Case A describes an unstrained system, whereas Case B (Case C) describes a system to which a uniaxial strain $\bm \sigma$($\perp$$\bm H$) ($\bm \sigma$($\parallel$$\bm H$)) is applied along the $x$ ($z$) axis. Indeed, our numerical calculations for Cases B and C reproduce reported phase diagrams obtained from experiments for $\bm \sigma$$\perp$$\bm H$ and $\bm \sigma$$\parallel$$\bm H$ in Refs.~\cite{Chacon15,Seki17}, respectively. For more quantitative discussions, it may be necessary to evaluate the strain-induced variation of the DM parameters microscopically using first-principles calculations~\cite{Koretsune15,Koretsune18}. However, this is beyond our present scope and is left for future studies.

We first investigate relative stabilities of various magnetic structures for Case A with $D_x$=$D_y$=$D_z$=0.727, Case B with $D_x$=0.8 and $D_y$=$D_z$=0.727, and Case C with $D_z$=0.8 and $D_x$=$D_y$=0.727. In Fig.~\ref{Fig1}(d)--(i), we show theoretical phase diagrams as function of $H$ at $T=0$ that were reproduced from calculated $H$-profiles of relative energies [Fig.~\ref{Fig1}(d)--(f)] and net magnetizations [Fig.~\ref{Fig1}(g)--(i)]. Here we examine five types of magnetic states: two different helical states (helical A and helical B) [Fig.~\ref{Fig1}(j) and (k)], a skyrmion crystal state [Fig.~\ref{Fig1}(l)], a conical state [Fig.~\ref{Fig1}(m)], and a ferromagnetic state. Note that helical A and helical B states have nearly the same energies, whereas their propagation vectors are slightly different. A slight difference in energy between these two different helical states might be an artifact of the finite-size effect. Specifically, in the present finite-sized cubic lattice, the helical state changes its propagation direction to fit its magnetic modulation period to the system size, which changes slightly upon the variation of $H$.

When the DM coupling is isotropic with $D_x$=$D_y$=$D_z$ as in Case A, only three magnetic phases, i.e., the helical A, conical, and ferromagnetic phases emerge successively as $H$ increases, and the skyrmion crystal phase does not appear [Fig.~\ref{Fig1}(d)]. This result is consistent with the fact that the skyrmion crystal phase appears only as a tiny pocket right below the magnetic transition temperature in the $T$--$H$ phase diagram for bulk chiral-lattice magnets.

In contrast, for Case B with $D_x$=0.8 and $D_y$=$D_z$=0.727, the skyrmion crystal phase appears sandwiched by the helical phase and the conical phase. Importantly, the regime for the conical phase is significantly suppressed, indicating that this state is destabilized by the $\bm H$ field applied perpendicular to the axial direction in which the DM interaction is stronger, and the skyrmion crystal phase attains a relative stability against the conical state. This result is consistent with the experimental observations that the skyrmion crystal phase is significantly stabilized and spreads even to the lowest temperature in the $T$--$H$ phase diagram when $\bm \sigma$$\perp$$\bm H$~\cite{Nii15,Chacon15,Seki17}.

For Case C with $D_z$=0.8 and $D_x$=$D_y$=0.727, the conical phase dominates the phase diagram, and the helical phase and the skyrmion crystal phase totally disappear. This is because the stronger DM coupling on bonds along the $z$ axis and the $\bm H$($\parallel$$\bm z$) field work cooperatively to stabilize the conical state. Specifically, the conical state propagating in the $z$ direction is characterized by the helically rotating magnetizations and the uniform component of magnetization along the $z$ axis, which are energetically favored by the strengthened DM interaction on bonds along the $z$ axis and the Zeeman interactions with $\bm H$($\parallel$$\bm z$). This result is again consistent with the experimental $T$--$H$ phase diagram with a dominant conical phase when $\bm \sigma$$\parallel$$\bm H$.

\begin{figure}[tb]
\includegraphics[scale=1.0]{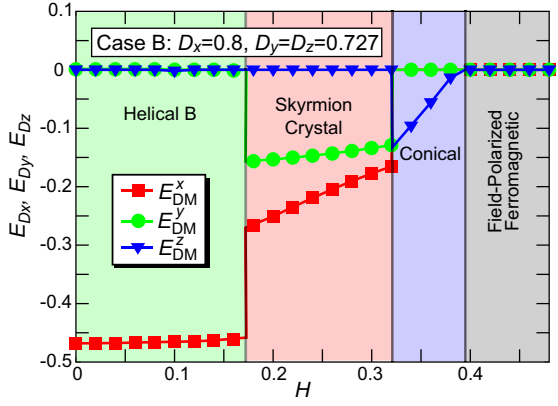}
\caption{Calculated $H$-profiles of energies $E_{\rm DM}^\alpha$ associated with the DM interactions on the bonds along the $\alpha$ axis ($\alpha=x, y, z$) at $T$=0 for Case B with $D_x$=0.8 and $D_y$=$D_z$=0.727.}
\label{Fig2}
\end{figure}
To get further insight into the strain-induced stabilization of magnetic phases, we calculated $H$-profiles of energies $E_{\rm DM}^\alpha$ associated with the DM parameters $D_\alpha$ on the bonds along the $\alpha$ axis ($\alpha$=$x$, $y$, $z$) at $T$=0. We numerically calculated them for Case B with $D_x$=0.8 and $D_y$=$D_z$=0.727 because all the relevant magnetic phases appear upon the variation of $H$ for this set of DM parameters. We find that the energy $E_{\rm DM}^x$ is negatively large in the helical phase, whereas the energies $E_{\rm DM}^x$ and $E_{\rm DM}^y$ are negatively large in the skyrmion crystal phase. On the contrary, the energy $E_{\rm DM}^z$ is almost zero in both the helical phase and the skyrmion crystal phase. These facts indicate that in these magnetic structures modulating within the $xy$ plane normal to the external magnetic field $\bm H$$\parallel$$\bm z$ is stabilized by the energy gain of DM interactions on the in-plane bonds characterized by the DM parameters $D_x$ and $D_y$. Thus, the increase of $D_x$ and/or $D_y$ under application of a uniaxial tensile strain $\sigma$ ($\perp$$\bm H$) energetically stabilizes them. On the other hand, the energy $E_{\rm DM}^z$ takes finite negative values in the conical phase, whereas the energies $E_{\rm DM}^x$ and $E_{\rm DM}^y$ are suppressed to be zero, indicating that the conical phase propagating along $\bm H$($\parallel$$\bm z$) is stabilized with increasing $D_z$ under application of uniaxial tensile strain $\sigma$ ($\parallel$$\bm H$).

In the present study, we examined the cases with $D_z>D_x=D_y$ and $D_x>D_y=D_z$ only because these conditions correspond to the situations of previous experimental studies which revealed the drastic strain-induced stabilization and/or destabilization of magnetic skyrmion phases in MnSi~\cite{Nii15,Chacon15} and Cu$_2$OSeO$_3$~\cite{Seki17}. On the other hand, we didn't study the cases with $D_z<D_x=D_y$ and $D_x<D_y=D_z$ although these conditions are also interesting to be examined. However, we can discuss what will happen for these cases on the basis of the above argument. The obtained $H$-profiles of partial DM energies indicate that the skyrmion crystal phase is stabilized in the case with $D_z<D_x=D_y$, whereas the helical phase propagating along the $y$ axis and the conical phase propagating along the $z$ axis are stabilized in the case with $D_x<D_y=D_z$. In the latter case, the skyrmion crystal phase will become unstable relative to the helical and conical phases.

\begin{figure}[tb]
\includegraphics[scale=1.0]{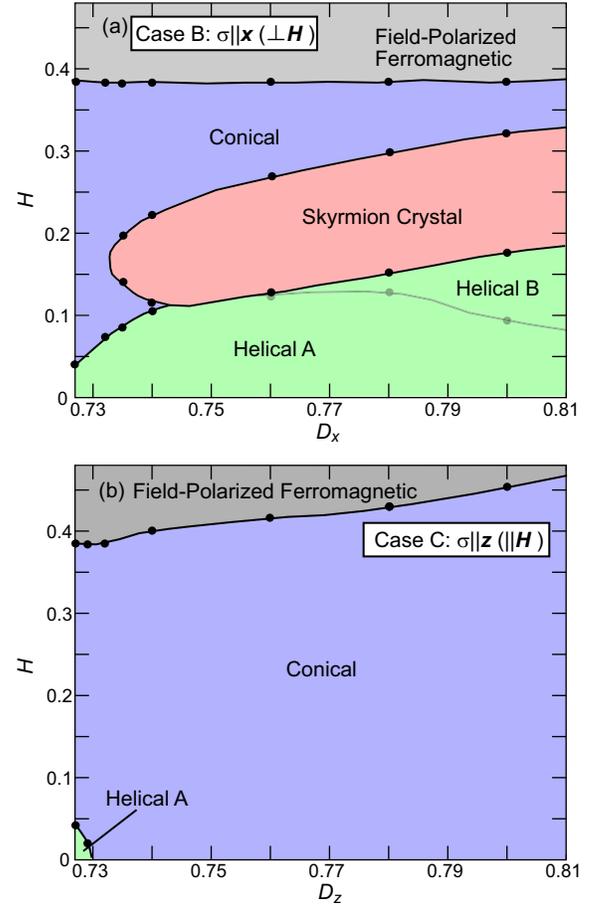}
\caption{(a) Theoretical phase diagram in the plane $D_x$ and $H$ for Case B with $D_x>D_y=D_z$ where $D_y$ and $D_z$ are fixed at 0.727. The condition corresponds to a system to which a uniaxial strain $\bm \sigma$($\parallel$$\bm x$) is applied perpendicular to $\bm H$($\perp$$\bm z$). (b) Theoretical phase diagram in the plane $D_z$ and $H$ for Case C with $D_z>D_x=D_y$ where $D_x$ and $D_y$ are fixed at 0.727. The condition corresponds to a system to which a uniaxial strain $\bm \sigma$($\parallel$$\bm z$) is applied parallel to $\bm H$($\parallel$$\bm z$).}
\label{Fig3}
\end{figure}
We next study the phase evolutions at $T$=0 with increasing anisotropy of the DM interactions. In Fig.~\ref{Fig3}(a), we show a phase diagram in the plane of $D_x$ and $H$ for Case B with $D_x>D_y=D_z$, where $D_y$ and $D_z$ are fixed at 0.727.
This condition corresponds to a system to which a uniaxial strain $\bm \sigma$($\parallel$$\bm x$) is applied perpendicular to $\bm H$($\perp$$\bm z$). The skyrmion crystal phase is absent when the DM coupling is isotropic with $D_x$=0.727, but it sets in above $D_x$$\sim$0.735. This indicates that only 1.1$\%$ anisotropy of the DM coupling perpendicular to $\bm H$ significantly stabilizes the skyrmion crystal phase. We also find that the helical phase is also enhanced by a tiny anisotropy of the DM coupling perpendicular to $\bm H$.

In contrast, we show a phase diagram in the plane of $D_z$ and $H$ for Case C with $D_z>D_x=D_y$ in Fig.~\ref{Fig3}(b), where $D_x$ and $D_y$ are fixed at 0.727. This condition corresponds to a system to which a uniaxial strain $\bm \sigma$($\parallel$$\bm z$) is applied parallel to $\bm H$($\perp$$\bm z$). Apparently, the phase diagram is dominated by the conical phase, whereas the helical phase, which exists when the DM coupling is isotropic when $D_z$=0.727, is abruptly suppressed as $D_z$ increases and disappears when $D_z$=0.73. This indicates that only a 0.4$\%$ anisotropy of the DM coupling parallel to $\bm H$ makes the conical state stable against other magnetic states.

\begin{figure}[tb]
\includegraphics[scale=1.0]{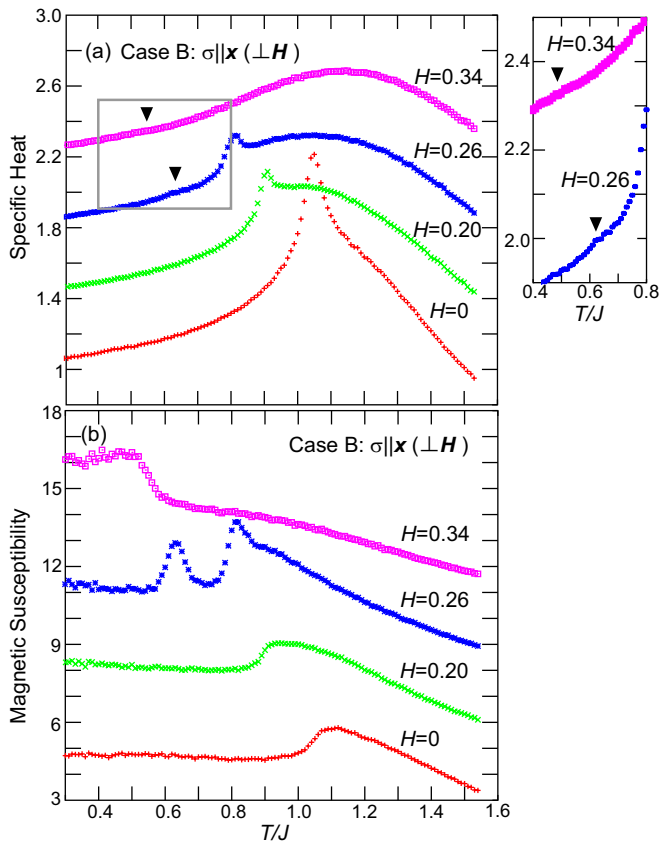}
\caption{Temperature profiles of (a) specific heats and (b) magnetic susceptibilities calculated using the replica-exchange Monte Carlo techniques for selected values of $H$ for Case B with $D_x$=0.8 and $D_y$=$D_z$=0.727, which corresponds to the condition $\sigma$$\perp$$\bm H$. Inverted triangles in (a) indicate transition points identified by the anomalies of the magnetic susceptibilities in (b) where no remarkable anomalies appear in the specific heats. The right panel of (a) magnifies the area indicated by the gray rectangle in the left panel of (a).}
\label{Fig4}
\end{figure}
We next study the thermodynamic properties of magnetic states at finite temperatures. For this purpose, we analyzed the classical spin model in Eq.~(\ref{eq:Htot}) using the replica-exchange Monte Carlo technique. We obtained $T$--$H$ phase diagrams by identifying phase-transition points from the calculated $T$-profiles of the specific heats and magnetic susceptibilities. Figures~\ref{Fig4}(a) and (b) show some examples of these $T$-profiles for selected values of $H$ when $D_x$=0.8 and $D_y$=$D_z$=0.727 (Case B).

At $H$=0 and $H$=0.2, the system exhibits a single phase transition to the helical phase and that to the skyrmion crystal phase, respectively, as temperature decreases. At the transition points, the specific heats exhibit a sharp peak, whereas the magnetic susceptibilities exhibit a kink with a sudden drop. The obtained $T$-profiles of specific heats and magnetic susceptibilities reproduce well the experimentally observed $T$-profiles of these quantities. At $H$=0.26, the system exhibits successive two phase transitions as temperature decreases. The system first enters the conical phase and subsequently the skyrmion crystal phase. The magnetic susceptibility shows peaks at the transition points, whereas we observe a prominent peak at the first transition but no remarkable anomaly at the second transition. Finally, at $H$=0.36, we again observe a single phase transition, at which the system enters the conical phase. At the transition point, the magnetic susceptibility shows a kink with a sudden rise, whereas the specific heat exhibits no remarkable anomaly. These behaviors are again in good agreement with experimental observations.

\begin{figure*}[t]
\includegraphics[scale=1.0]{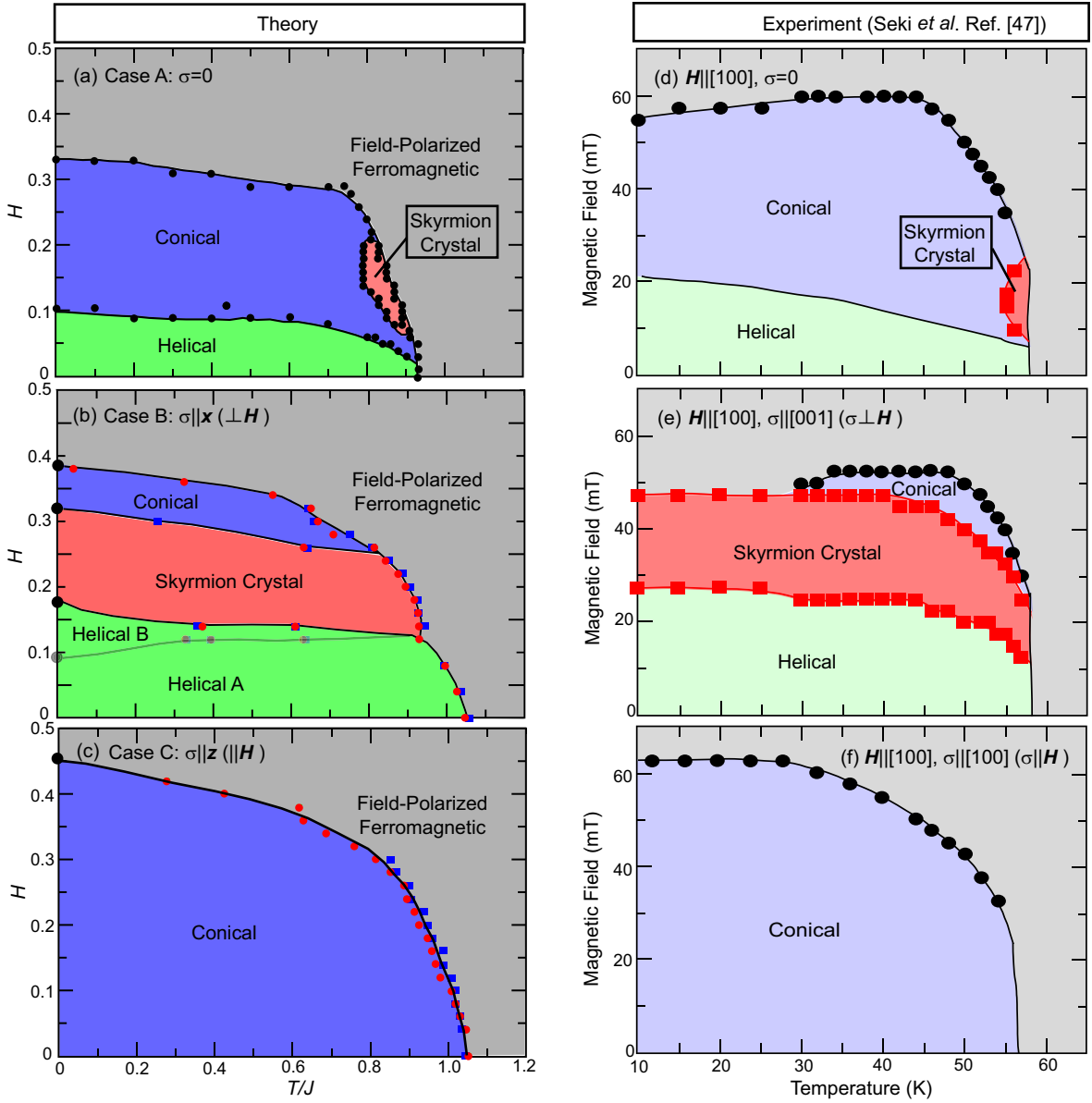}
\caption{(a) Theoretical $T$--$H$ phase diagram for Case A with $D_x$=$D_y$=$D_z$=0.727, which corresponds to an unstrained system with isotropic DM coupling (Reproduced from Ref.~\cite{Buhrandt13}). (b) Theoretical $T$--$H$ phase diagram for Case B with $D_x$=0.8 and $D_y$=$D_z$=0.727, which corresponds to a system to which a uniaxial strain $\bm \sigma$($\perp$$\bm H$) is applied. (c) Theoretical $T$--$H$ phase diagram for Case C with $D_z$=0.8 and $D_x$=$D_y$=0.727, which corresponds to a system to which a uniaxial strain $\bm \sigma$($\parallel$$\bm H$) is applied. Circles and squares indicate transition points identified by anomalies in specific heats and magnetic susceptibilities, respectively. (d) Experimental $T$--$H$ phase diagram for Cu$_2$OSeO$_3$ without strain. (e) Experimental $T$--$H$ phase diagram for Cu$_2$OSeO$_3$ under $\bm H$$\parallel$$[100]$ to which a uniaxial strain $\bm \sigma$$\parallel$$[001]$ ($\perp$$\bm H$) is applied. (f) Experimental $T$--$H$ phase diagram for Cu$_2$OSeO$_3$ under $\bm H$$\parallel$$[100]$ to which a uniaxial strain $\bm \sigma$$\parallel$$[100]$ ($\parallel$$\bm H$) is applied. The experimental phase diagrams in (d)-(f) are reproduced from Ref.~\cite{Seki17}.}
\label{Fig5}
\end{figure*}
In Fig.~\ref{Fig5}(a)-(c), we display three theoretical phase diagrams in the plane of $T$ and $H$, which were calculated for Case A with $D_x$=$D_y$=$D_z$=0.727, Case B with $D_x$=0.8 and $D_y$=$D_z$=0.727, and Case C with $D_z$=0.8 and $D_x$=$D_y$=0.727. The phase diagram in Fig.~\ref{Fig5}(a) for the isotropic DM interactions is reproduced from previous theoretical work by Buhrandt and Fritz~\cite{Buhrandt13}, whereas those in Fig.~\ref{Fig5}(b) and (c) were obtained in the present work. We also display the experimental $T$--$H$ phase diagrams for Cu$_2$OSeO$_3$ under application of magnetic field $\bm H$$\parallel$$[100]$ in Fig.~\ref{Fig5}(d)--(f), which are reproduced from Ref.~\cite{Seki17}. The phase diagram in Fig.~\ref{Fig5}(d) is obtained for an unstrained sample, whereas the phase diagrams in Fig.~\ref{Fig5}(e) and (f) are obtained for strained samples.

When the DM coupling is isotropic, as in Case A, the skyrmion crystal phase appears as a tiny pocket on the verge of the phase boundary between the paramagnetic phase and the conical phase [Fig.~\ref{Fig5}(a)] in agreement with the experimental observations [see Fig.~\ref{Fig5}(d)]. This situation changes enormously when we introduce the uniaxial anisotropy of DM interactions by applying a uniaxial strain. When the DM coupling is strengthened on bonds perpendicular to $\bm H$, as in Case B, the skyrmion crystal phase is significantly stabilized and spreads even into the low temperatures [Fig.~\ref{Fig5}(b)]. This result reproduces well the experimental phase diagram obtained for $\bm \sigma$$\perp$$\bm H$ in Fig.~\ref{Fig5}(e). In contrast, when the DM coupling is strengthened on bonds parallel to $\bm H$ as in Case C, the skyrmion crystal phase vanishes, and the $T$--$H$ phase diagram is dominated by the conical phase propagating in the $\bm H$ direction. This result again reproduces well the experimental phase diagram obtained for $\bm \sigma$$\parallel$$\bm H$ in Fig.~\ref{Fig5}(f).

\section{Conclusion}
Motivated by recent experimental findings of strain-induced stabilization and destabilization of magnetic skyrmions in bulk chiral-lattice ferromagnets MnSi~\cite{Nii15,Chacon15} and Cu$_2$OSeO$_3$~\cite{Seki17}, we studied the effects of anisotropic DM interactions on the stability of the skyrmion crystal phase in a numerical analysis of the classical spin model. We found that the anisotropic DM interactions significantly enhance or suppress the stability of skyrmion crystal phase depending on the relative direction of their anisotropy axis against the external $\bm H$ field. More specifically, when the DM coupling perpendicular (parallel) to $\bm H$ is strengthened, the skyrmion crystal phase is stabilized (destabilized). Our Monte Carlo calculations reproduced the experimentally observed $T$--$H$ phase diagrams for both $\bm \sigma$$\perp$$\bm H$ and $\bm \sigma$$\parallel$$\bm H$. Our results support that the application of uniaxial strain indeed controls the stability of magnetic skyrmions via inducing the anisotropic DM coupling and thus provide firm ground for possible strain engineering of magnetic skyrmions towards future skyrmion-based electronics.

\section{Acknowledgment}
We thank Y. Takahashi for her technical help in the data analyses. This work was supported by JSPS KAKENHI (Grant Nos. 17H02924, 16H06345, 19H00864, and 19K21858) and Waseda University Grant for Special Research Projects (Project No. 2019C-253).


\begin{thebibliography}{999}
\bibitem{Dzyaloshinsky58}I. Dzyaloshinsky, J. Phys. Chem. Solids {\bf 4}, 241 (1958).

\bibitem{Moriya60}T. Moriya, Phys. Rev. {\bf 120}, 91 (1960).

\bibitem{Bogdanov89}A. N. Bogdanov and D. A. Yablonskii, Sov. Phys. JETP {\bf 68}, 101 (1989).

\bibitem{Bogdanov99}A. Bogdanov and H. Hubert, J. Mag. Mag. Mat. {\bf 195}, 182 (1999).


\bibitem{Rossler06}U. K. R\"osler, A. N. Bogdanov, and C. Pfleiderer, Nature (London) {\bf 442}, 797 (2006).

\bibitem{Nagaosa13}N. Nagaosa and Y. Tokura, Nat. Nanotech. {\bf 8}, 899 (2013).


\bibitem{Everschor18}K. Everschor-Sitte, J. Masell, R. M. Reeve, and M. Kl\"aui, J. Appl. Phys. {\bf 124}, 240901 (2018).

\bibitem{Muhlbauer09}S. M\"uhlbauer, B. Binz, F. Jonietz, C. Pfleiderer, A. Rosch, A. Neubauer, R. Georgii, and P. B\"oni, Science {\bf 323}, 915 (2009).

\bibitem{Tonomura12}A. Tonomura, X. Z. Yu, K. Yanagisawa, T. Matsuda, Y. Onose, N. Kanazawa, H. S. Park, and Y. Tokura, Nano Lett. {\bf 12}, 1673 (2012).

\bibitem{YuXZ10}X. Z. Yu, Y. Onose, N. Kanazawa, J. H. Park, J. H. Han, Y. Matsui, N. Nagaosa, and Y. Tokura, Nature (London) {\bf 465}, 901 (2010).

\bibitem{Munzer10}W. M\"unzer, A. Neubauer, T. Adams, S. M\"uhlbauer, C. Franz, F. Jonietz, R. Georgii, P. B\"oni, B. Pedersen, M. Schmidt, A. Rosch, and C. Pfleiderer, Phys. Rev. B, {\bf 81}, 041203(R) (2010).

\bibitem{Jonietz10}F. Jonietz, S. M\"uhlbauer, C. Pfleiderer,A. Neubauer,W. M\"unzer, A. Bauer, T. Adams, R. Georgii, P. Boni, R. A. Duine, K. Everschor, M. Garst, and A. Rosch, Science {\bf 330}, 1648 (2010).

\bibitem{YuXZ12}X. Z. Yu, N. Kanazawa, W. Z. Zhang, T. Nagai, T. Hara, K. Kimoto, Y. Matsui, Y. Onose, and Y. Tokura, Nat. Commun. {\bf 3}, 988 (2012).

\bibitem{Schulz12}T. Schulz, R. Ritz, A. Bauer, M. Halder, M. Wagner, C. Franz, C. Pfleiderer, K. Everschor, M. Garst, and A. Rosch, Nat. Phys. {\bf 8}, 301 (2012).

\bibitem{Yamanouchi04}M. Yamanouchi, D. Chiba, F. Matsukura, and H. Ohno, Nature, (London) {\bf 428}, 539 (2004).

\bibitem{Tatara04}G. Tatara, and H. Kohno, Phys. Rev. Lett. {\bf 92}, 086601 (2004).

\bibitem{Barnes05}S. E. Barnes, and S. Maekawa, Phys. Rev. Lett. {\bf 95}, 107204 (2005).

\bibitem{Everschor12}K. Everschor, M. Garst, B. Binz, F. Jonietz, S. M\"uhlbauer, C. Pfleiderer, and A. Rosch, Phys. Rev. B {\bf 86}, 054432 (2012).

\bibitem{Iwasaki13a}J. Iwasaki, M. Mochizuki, and N. Nagaosa, Nat. Commun. {\bf 4}, 1463 (2013).

\bibitem{Iwasaki13b}J. Iwasaki, M. Mochizuki, and N. Nagaosa, Nat. Nanotech. {\bf 8}, 742 (2013).

\bibitem{Fert13}A. Fert, V. Cros, and J. Sampaio, Nat. Nanotech. {\bf 8}, 152 (2013).

\bibitem{Tomasello14}R. Tomasello, E. Martinez, R. Zivieri, L. Torres, M. Carpentieri, and G. Finocchio, Sci. Rep. {\bf 4}, 6784 (2014).

\bibitem{Koshibae15} W. Koshibae, Y. Kaneko, J. Iwasaki, M. Kawasaki, Y. Tokura, and N. Nagaosa, Jpn. J. Appl. Phys. {\bf 54}, 053001 (2015).

\bibitem{Finocchio16}G. Finocchio, F. B\"uttner, R. Tomasello, M. Carpentieri, and M. Kl\"aui, J. Phys. D: Appl. Phys. {\bf 49}, 423001 (2016).

\bibitem{ZhangX15}X. Zhang, M. Ezawa, and Y. Zhou, Sci. Rep. {\bf 5}, 9400 (2015).

\bibitem{Finocchio15}G. Finocchio, M. Ricci, R. Tomasello, A. Giordano, M. Lanuzza, V. Puliafito, P. Burrascano, B. Azzerboni, and M. Carpentieri, Appl. Phys. Lett. {\bf 107}, 262401 (2015).



\bibitem{Pinna18b}D. Pinna, F. Abreu Araujo, J.-V. Kim, V. Cros, D. Querlioz, P. Bessiere, J. Droulez, and J. Grollier, Phys. Rev. Appl. {\bf 9}, 064018 (2018).

\bibitem{Prychynenko18}D. Prychynenko, M. Sitte, K. Litzius, B. Kr\"uger, G. Bourianoff, M. Kl\"aui, J. Sinova, and K. Everschor-Sitte, Phys. Rev. Appl. {\bf 9}, 014034 (2018).

\bibitem{Bourianoff18}G. Bourianoff, D. Pinna, M. Sitte, and K. Everschor-Sitte, AIP Advances {\bf 8}, 055602 (2018).

\bibitem{Neubauer09}A. Neubauer, C. Pfleiderer, B. Binz, A. Rosch, R. Ritz, P. G. Niklowitz, and P. Boni, Phys. Rev. Lett. {\bf 102}, 186602 (2009).

\bibitem{Wilhelm11}H. Wilhelm, M. Baenitz, M. Schmidt, U. K. R\"ossler, A. A. Leonov, and A. N. Bogdanov, Phys. Rev. Lett. {\bf 107}, 127203 (2011).

\bibitem{Seki12a}S. Seki, S. Ishiwata, and Y. Tokura, Science, {\bf 336}, 198 (2012).

\bibitem{Seki12b}S. Seki, S. Ishiwata, and Y. Tokura, Phys. Rev. B {\bf 86}, 060403(R) (2012).

\bibitem{Seki12c}S. Seki, J.-H. Kim, D. S. Inosov, R. Georgii, B. Keimer, S. Ishiwata, and Y. Tokura, Phys. Rev. B {\bf 85}, 220406(R) (2012).

\bibitem{Adams12}T. Adams, A. Chacon, M. Wagner, A. Bauer, G. Brandl, B. Pedersen, H. Berger, P. Lemmens, and C. Pfleiderer, Phys. Rev. Lett. {\bf 108}, 237204 (2012).

\bibitem{Kezsmarki15}I. Kezsmarki, $et\ al$. Nat. Mater. {\bf 14}, 1116 (2015).

\bibitem{Tokunaga15}Y. Tokunaga, $et\ al$., Nat. Commun. {\bf 6}, 7638 (2015).

\bibitem{YuXZ11}X. Z. Yu, N. Kanazawa, Y. Onose, K. Kimoto, W. Z. Zhang, S. Ishiwata, Y. Matsui, and Y. Tokura, Nat. Mater. {\bf 10}, 106 (2011).

\bibitem{Butenko10}A. B. Butenko, A. A. Leonov, U. K. R\"ossler, and A. N. Bogdanov, Phys. Rev. B {\bf 82}, 052403 (2010).

\bibitem{Ritz13}R. Ritz, M. Halder, M. Wagner, C. Franz, A. Bauer, and C. Pfleiderer, Nature (London) {\bf 497}, 231 (2013).

\bibitem{Levatic16}I. Levati\'{c}, P. Popevi\'{c}, V. \v{S}urija, A. Kruchkov, H. Berger, A. Magrez, J. S. White, H. M. R{\o}nnow, and I. \v{Z}ivkovi\'{c}, Sci. Rep. {\bf 6}, 21347 (2016).

\bibitem{Okamura16}Y. Okamura, F. Kagawa, S. Seki, and Y. Tokura, Nat. Commun. {\bf 7}, 12669 (2016).

\bibitem{Oike16}H. Oike, A. Kikkawa, N. Kanazawa, Y. Taguchi, M. Kawasaki, Y. Tokura, and F. Kagawa, Nat. Phys. {\bf 12}, 62 (2016).

\bibitem{Karube16}K. Karube, J. S. White, N. Reynolds, J. L. Gavilano, H. Oike, A. Kikkawa, F. Kagawa, Y. Tokunaga, H. M. R{\o}nnow, Y. Tokura, and Y. Taguchi, Nat. Mater. {\bf 15}, 1237 (2016).

\bibitem{Nii15}Y. Nii, T. Nakajima, A. Kikkawa, Y. Yamasaki, K. Ohishi, J. Suzuki, Y. Taguchi, T. Arima, Y. Tokura, and Y. Iwasa, Nat. Commun. {\bf 6}, 8539 (2015).

\bibitem{Chacon15}A. Chacon, A. Bauer, T. Adams, F. Rucker, G. Brandl, R. Georgii, M. Garst, and C. Pfleiderer, Phys. Rev. Lett. {\bf 115}, 267202 (2015).

\bibitem{Seki17}S. Seki, Y. Okamura, K. Shibata, R. Takagi, N. D. Khanh, F. Kagawa, T. Arima, and Y. Tokura, Phys. Rev. B {\bf 96}, 220404(R) (2017).

\bibitem{Bak80}P. Bak and M. H. Jensen, J. Phys. C: Solid State Phys. {\bf 13}, L881 (1980).

\bibitem{Buhrandt13}S. Buhrandt and L. Fritz, Phys. Rev. B, {\bf 88}, 195137 (2013).

\bibitem{Swendsen86}R. H. Swendsen and J. S. Wang, Phys. Rev. Lett. {\bf 57}, 2607 (1986).

\bibitem{Hukushima96}K. Hukushima and K. Nemoto, J. Phys. Soc. Jan. {\bf 65}, 1604 (1996).

\bibitem{Earl05}D. J. Earl and M. W. Deem, Phys. Chem. Chem. Phys. {\bf 7}, 3910 (2005).

\bibitem{Shibata15}K. Shibata, J. Iwasaki, N. Kanazawa, S. Aizawa, T. Tanigaki, M. Shirai, T. Nakajima, M. Kubota, M. Kawasaki, H. S. Park, D. Shindo, N. Nagaosa, and Y. Tokura, Nat. Nanotech. {\bf 10}, 589 (2015).

\bibitem{Koretsune15}T. Koretsune, N. Nagaosa, and R. Arita, Sci. Rep. {\bf 5}, 13302 (2015).

\bibitem{Koretsune18}T. Koretsune, T. Kikuchi, and R. Arita, J. Phys. Soc. Jpn. {\bf 87}, 041011 (2018).
\end{thebibliography}
\end{document}